\documentclass[fleqn,usenatbib]{mnras}

\usepackage{newtxtext,newtxmath}
\usepackage[T1]{fontenc}
\DeclareRobustCommand{\VAN}[3]{#2}
\let\VANthebibliography\thebibliography
\def\thebibliography{\DeclareRobustCommand{\VAN}[3]{##3}\VANthebibliography}

\usepackage{graphicx}	
\usepackage{amsmath}	
\usepackage{orcidlink}
\usepackage{multirow}
\usepackage{subcaption}

\defcitealias{2007MNRAS.381..543W}{W07}
\defcitealias{2013ApJ...765...28A}{A13}
\defcitealias{2022MNRAS.509.3966W}{Wa22}
\defcitealias{2022MNRAS.516.4354W}{Wi22}
\defcitealias{2022MNRAS.517L..92E}{E22}
\defcitealias{2018MNRAS.477.1708P}{P18}
\defcitealias{2024MNRAS.534.1459G}{G24}

\title[Enhanced star formation and tidal features]{Linking enhanced star formation and quenching to faint tidal features in galaxies}

\author[A. J. Gordon et al.]{
Alexander J. Gordon$^{1}$\thanks{E-mail: alexander.gordon@ed.ac.uk}\orcidlink{0009-0006-4035-5019},
Annette M. N. Ferguson$^{1}$,
Robert G. Mann$^{1}$\orcidlink{0000-0002-0194-325X}
and Vivienne Wild$^{2}$\orcidlink{0000-0002-8956-7024}
\\
$^{1}$ Institute for Astronomy, University of Edinburgh, Royal Observatory, Blackford Hill, Edinburgh EH9 3HJ, UK\\
$^{2}$ School of Physics \& Astronomy, University of St Andrews, North Haugh, St Andrews, KY16 9SS, UK\\
}

\date{Accepted XXX. Received YYY; in original form ZZZ}

\pubyear{2025}

\begin{document}
\label{firstpage}
\pagerange{\pageref{firstpage}--\pageref{lastpage}}
\maketitle

\begin{abstract}
Galaxy mergers and interactions have long been suggested as a significant driver of galaxy evolution. However, the exact extent to which mergers enhance star formation and AGN activity has been challenging to establish observationally. In previous work, we visually classified a sample of galaxies with various types of faint tidal features in DECaLS images. In this paper, we cross-correlate this sample with SDSS-derived data to investigate how the presence and specific nature of these features correlates with intense star formation and AGN activity. Averaged over all tidal classes, we find that our 688 tidal feature galaxies are 6.6$\pm$0.9 times more likely to be in a starburst phase and 19.6$\pm$5.0 times more likely to have rapidly quenched (post-starbursts) than a sample of 4073 controls matched in both stellar mass and redshift. Examining differences between tidal classes, galaxies with \textit{arm} features were $\sim$1.3-4.0 times more likely to be starbursting than the other categories, while those with \textit{shell} features were $\sim$2.3-5.3 times more likely to be in a quiescent state. In a similar analysis, we identify which galaxies show evidence of AGN activity (from a sample of $\sim$2100) and find no significant difference between those with or without tidal features. Overall, our results reinforce the notion that mergers play an important role in driving star formation and rapid quenching in galaxies, and provide some of the first empirical evidence that the strength of this effect has a dependence on the detailed nature of the interaction, as traced by the tidal feature morphology.
\end{abstract}




\section{Introduction}
The hierarchical growth of galaxies is driven by interactions and mergers between galaxies \citep[e.g.][]{1978MNRAS.183..341W, 1991ApJ...379...52W}. Major (mass ratio $\mu\sim0.25-1.0$), minor ($\mu\sim0.1-0.25$), and mini \citep[$\mu\lesssim0.1$, see e.g.][]{2024MNRAS.527.6506B} mergers can all contribute a considerable proportion of the mass in a galaxy \citep[e.g.][]{2010ApJ...725.2312O, 2016MNRAS.458.2371R}. However, as minor and mini mergers occur more frequently than majors \citep[e.g.][]{2010MNRAS.406.2267F}, they may be the primary way in which mergers contribute to mass build-up \citep[e.g.][]{2014MNRAS.445.2198O, 2016ApJ...821....5D, 2024MNRAS.527.6506B}. All these types of interactions can leave behind debris \citep[e.g.][]{1972ApJ...178..623T, 2010MNRAS.406..744C}, either from recently accreted material or late-stage relics, and are known as tidal features.

Simulations suggest that during mergers and interactions, tidal forces generate torques that lead to the inflow of gas \citep[e.g.][]{1996ApJ...471..115B, 1996ApJ...464..641M}. Furthermore, this inflow of gas, as well as tidal compression and shocks, can trigger intense star formation in the nucleus of a galaxy \citep[starbursts; e.g.][]{1994ApJ...425L..13M,2007A&A...468...61D,2015MNRAS.448.1107M}. This has been supported by observations that have found enhancements in star formation in close pairs of galaxies, in late-stage mergers, and in coalesced post-merger systems, with the most significant enhancements occurring at low projected separations and coalesced systems \citep[e.g.][]{2008AJ....135.1877E, 2011MNRAS.412..591P, 2012MNRAS.426..549S, 2013MNRAS.435.3627E, 2014AJ....148..137L, 2025MNRAS.538L..31F}. However, 
many details of the nature of merger-driven star formation remain unclear. For example,  while some studies find that the star formation rate (SFR) is boosted in merging systems by a factor of between $\sim$1.2 and $\sim$3.5 compared to controls \citep[e.g.][]{2012MNRAS.426..549S,2013MNRAS.435.3627E,2015MNRAS.454.1742K,2019A&A...631A..51P}, others have suggested that certain merger events can lead to no increase or even a decrease in the SFR \citep[e.g.][]{2015MNRAS.454.1742K, 2019A&A...631A..51P, 2023ApJ...953...91L, 2025ApJ...979....7L}. Another open question concerns the minimum merger mass ratio necessary to lead to significant enhanced star formation \citep[e.g.][]{2008MNRAS.384..386C}. 

Mergers have also been implicated in the rapid quenching of starbursts and their evolution into the post-starburst phase, but the picture is not fully complete. While the observed fraction of post-starbursts exhibiting evidence of recent mergers is significantly higher than those not showing signs of interaction \citep[e.g.][]{2018MNRAS.477.1708P,2021ApJ...919..134S,2022MNRAS.516.4354W,2023ApJ...949....5V},  some simulation studies find no link \citep[e.g.][]{2019MNRAS.490.2139R}, while others only see a link in major mergers with specific orbital parameters \citep[e.g.][]{2020MNRAS.498.1259Z}. \citet[][hereinafter \citetalias{2022MNRAS.517L..92E}]{2022MNRAS.517L..92E} build upon the work of \citet[][hereinafter \citetalias{2018MNRAS.477.1708P}]{2018MNRAS.477.1708P} and \citet[][hereinafter \citetalias{2022MNRAS.516.4354W}]{2022MNRAS.516.4354W} and address the reverse problem, namely what fraction of post-merger systems have recently quenched their star formation. They find a frequency of post-starbursts that is between 30 and 60 times greater than that of the non-merger controls. Furthermore, they find that this excess does not occur in a sample of close pairs; thus, they argue that mergers can cause the rapid quenching of star formation, but only after coalescence.

Simulations have also suggested that the merger-induced inflow of gas can trigger active galactic nuclei (AGN\footnote{In this work we will use AGN to refer to both the AGN itself and the AGN host.}) by increasing the accretion of material onto the central supermassive black hole \citep[e.g.][]{2000MNRAS.311..576K, 2005MNRAS.361..776S, 2018MNRAS.479.3952B, 2024MNRAS.528.5864B}. However, this has been more challenging to demonstrate observationally. Several studies support the idea that the mergers and AGN are linked by studying AGN in close pairs \citep[e.g.][]{1985AJ.....90..708K, 2007MNRAS.375.1017A, 2007AJ....134..527W}, while others find an enhanced fraction of AGN showing disturbed morphologies compared to controls \citep[e.g.][]{2012MNRAS.426..276B, 2014MNRAS.441.1297S, 2018PASJ...70S..37G}. On the other hand, several studies have failed to find any convincing link between mergers and AGN activity \citep[e.g.][]{2009ApJ...691.1005R, 2016ApJ...830..156M, 2019MNRAS.483.2441V, 2024A&A...691A..82C}. There are numerous possible reasons why these studies may have differing results, including sample biases and selection effects \citep[see, e.g.][]{2023OJAp....6E..34V, 2025OJAp....8E..12E}. Most of the works mentioned above begin by identifying AGN using specific criteria and then determining the fraction of those that exhibit disturbed morphologies. In particular, the focus is often on clear merging signatures, such as pairs and bright tidal features, the visibility of which depends on survey depth \citep[e.g.][]{2008AJ....135.1877E}. 

Indeed, it is well established that signatures of past mergers and interactions become increasingly common as fainter surface brightness depths are probed 
\citep[e.g.][]{2008ApJ...689..936J, 2022MNRAS.513.1459M}. In the very local Universe, where star count techniques can be used to probe to extremely low surface brightnesses ($\gtrsim 30$ mag arcsec$^{-2}$), galaxies that appear completely undisturbed when viewed at high surface brightness are often revealed to be surrounded by networks of faint tidal features in deep images \citep[e.g.][]{2002AJ1241452F, 2015ApJ809L1O, 2025ApJ...982L..41F}. This faint debris can have a variety of origins, including recent minor or mini-merger events \citep[e.g.][]{2008ApJ...689..936J, 2019MNRAS.487..318K}, major mergers which occurred many gigayears ago \citep[e.g.][]{1972ApJ...178..623T}, or recent fly-by interactions \citep[e.g.][]{2014ApJ...789...90K}. With new and forthcoming large imaging surveys, the links between galaxy interactions and host galaxy properties can be probed in a statistical sense to much fainter surface brightnesses than ever before, providing sensitivity to a much broader range of merger and interaction events. With such data, one can not only address the question of how mergers affect SFRs and nuclear activity in a general sense, but also how effective specific types of merger events are in triggering, or quenching, this activity. In a similar vein, recent work has explored the links between different types of tidal features and the kinematical properties of galaxies, finding that galaxies with shells typically rotate more slowly than galaxies with streams \citep[e.g.][]{2024A&A...686A.182V, 2024ApJ...965..158Y}. Since shells are commonly associated with radial mergers, this suggests that such events are more efficient in decreasing a galaxy's angular momentum than mergers with more circular orbits, which lead to stream formation.

In \citet[][hereinafter \citetalias{2024MNRAS.534.1459G}]{2024MNRAS.534.1459G} we identified a sample of $\sim$950 galaxies with tidal features in Dark Energy Camera Legacy Survey \citep[DECaLS;][]{2019AJ....157..168D} DR5 data and categorised them into four different morphologies. In this work, we exploit this sample to undertake the first investigation of how star formation and AGN activity vary across galaxies exhibiting different tidal feature classes, as well as when compared to non-merging controls. The paper is structured as follows. Section \ref{sec:data} details the construction of the galaxy sample and how these were identified as having tidal features. Section \ref{sec:selection} describes our stellar mass and redshift cuts, how these were matched between the galaxies with tidal features, and how we identified the evolutionary phase and AGN activity of each galaxy. Section \ref{sec:results} presents the results of this work, and some discussion is presented in Section \ref{sec:discussion}, with Section \ref{sec:summary} providing a summary.

\section{Data and morphology}\label{sec:data}
\subsection{DECaLS sample}\label{sec:decalsmorph}
The primary focus of DECaLS was to identify targets for the Dark Energy Spectroscopic Instrument (DESI) survey. Using the 4 m Blanco telescope at the Cerro Tololo Inter-American Observatory in Chile, it imaged around 9000 deg$^2$ of the sky in the \textit{g}, \textit{r}, and \textit{z} bands. We followed the standard method of estimating the limiting surface brightness depth of the data using 3$\sigma$ of the sky in $10 \times 10$ arcsec$^2$ boxes \citep{romn2020galactic-27e}, and found an average value of $\mu_r \sim$28.0\footnote{This is fainter than the value reported in \citepalias{2024MNRAS.534.1459G} due to a calculation error.} mag arcsec$^{-2}$ across our sample.  However, individual galaxy cut-outs had limiting surface brightness depths considerably higher or lower than this, with measured values ranging from $\sim 27.2-29.2$ mag arcsec$^{-2}$.

\begin{figure}
    \includegraphics[width=\columnwidth]{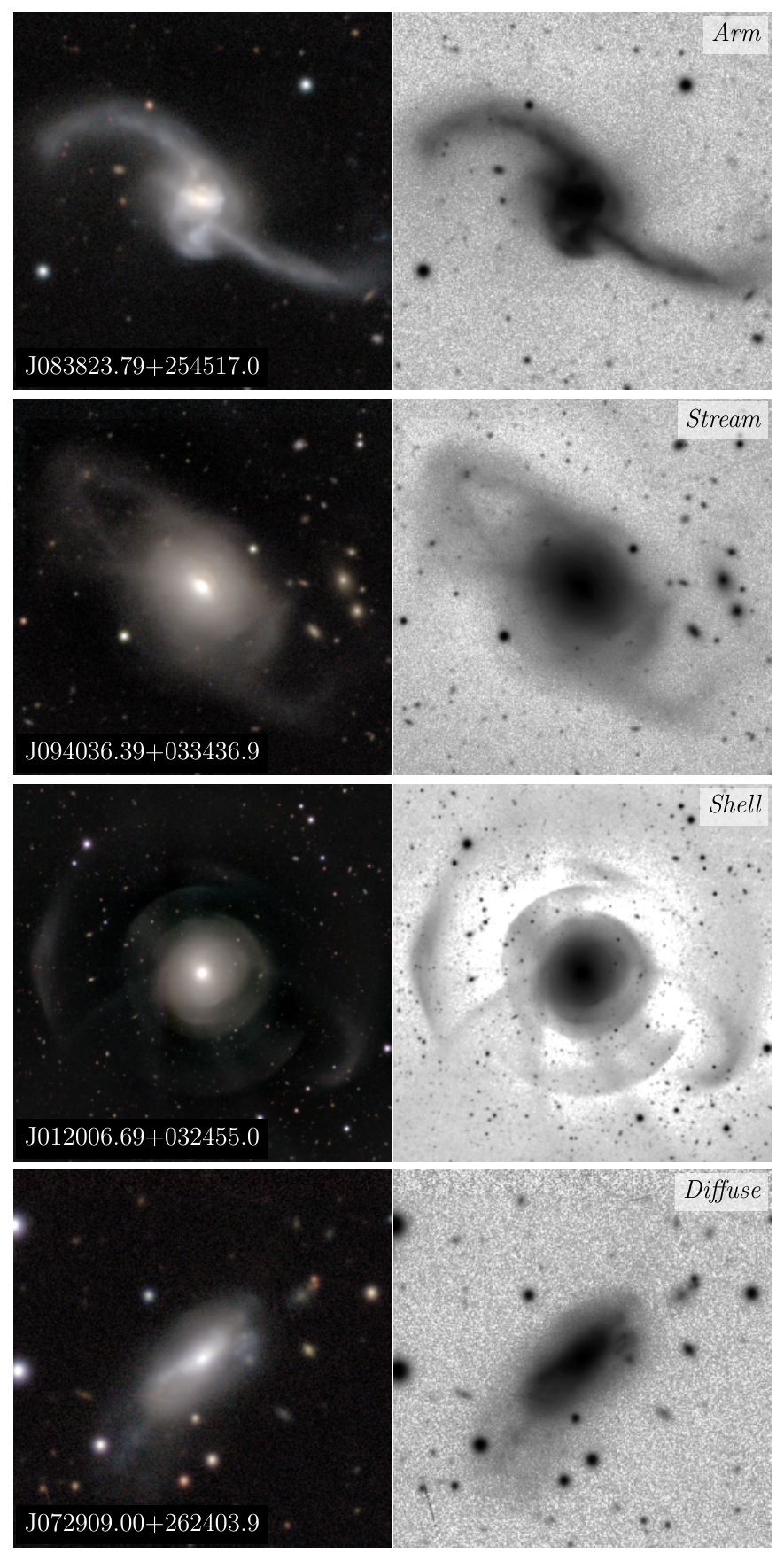}
    \caption{Example thumbnails of the four categories of tidal features used in this work: \textit{arms} (top), \textit{streams} (middle-top), \textit{shells} (middle-bottom), and \textit{diffuse} (bottom). The images on the left were created by \citetalias{2022MNRAS.509.3966W} and those on the right were created in \citetalias{2024MNRAS.534.1459G}.}
    \label{fig:example_tidal_features}
\end{figure}

As mentioned, we used the sample of galaxies with tidal features (hereinafter tidal galaxies) previously identified in DECaLS DR5 data, along with a corresponding control sample. A summary of the process used to identify these features is as follows \citepalias[see][for full details]{2024MNRAS.534.1459G}. A sample of candidate galaxies likely to host tidal features was determined from the \citet[hereinafter \citetalias{2022MNRAS.509.3966W}]{2022MNRAS.509.3966W} catalogue of bulk morphology predictions\footnote{\url{https://zenodo.org/records/4573248} -- gz\_decals\_auto\_posteriors}. The magnitudes and redshifts of the galaxies were limited to $-19\geq M_r\geq -22$ and $z\leq0.15$, from their values in the NASA Sloan Atlas\footnote{v1\_0\_1 available at: \url{https://www.sdss4.org/dr17/manga/manga-target-selection/nsa/}} \citep[NSA;][]{2011AJ....142...31B}. Galaxies indicated in \citetalias{2022MNRAS.509.3966W} as potentially having artefacts were removed\footnote{\texttt{artifact\_fraction} $\leq$ 0.1}. The sample was then split into candidates and controls based on the \texttt{merging} predictions; those with \verb|merging_minor| or \verb|merging_major| predictions greater than 0.4 were taken as candidates, and those with \verb|merging_minor|, \verb|merging_major|, and \verb|merging_merger| all less than 0.08 were taken as controls. Each of the candidates was then visually inspected to identify which categories of tidal feature it had. The four non-exclusive options were: \textit{arm}, \textit{stream}, \textit{shell}, and \textit{diffuse}. These categories were motivated by the appearance of the features \citepalias[see][for a full description of these]{2024MNRAS.534.1459G} and an example of each is provided in Fig. \ref{fig:example_tidal_features}. \textit{Shells} exhibited some symmetry and included well-defined or brighter edges, such as a fan shape or concentric arcs. \textit{Arms} generally were connected to the host, were fairly broad in width, and the surface brightness tailed off with increasing distance from the host. \textit{Streams} had similar shapes to \textit{arms}; however, they did not appear smoothly connected to the host, were much narrower and often had peak surface brightness far from the host. \textit{Diffuse} captures the class where the feature appeared irregular or asymmetric, or where it did not reasonably resemble one of the other classes. In total there were around 950 galaxies identified in \citetalias{2024MNRAS.534.1459G} as having tidal features.

In this work, we expanded the control sample to include a larger number of galaxies by loosening the restriction on the \verb|merging_minor|, \verb|merging_major|, and \verb|merging_merger| columns, where each had to be less than 0.15. This allowed us to generate a pool of $\sim$80$\,$000 potential controls that we later cross-matched to SDSS data and ensured that the stellar mass and redshift distributions matched that of the tidal galaxies (see Section \ref{sec:massredshift}). The magnitude and artefact cuts were kept as they were in \citetalias{2024MNRAS.534.1459G}. We further categorised the galaxies in this control pool into distinct morphologies by referring back to the predictions of \citetalias{2022MNRAS.509.3966W}. Based on these predictions, we created four categories for the control galaxies: elliptical, spiral, edge-on disc, and miscellaneous disc. Following the cuts suggested by \citetalias{2022MNRAS.509.3966W}, we took any galaxy with a prediction greater than 0.7 for \verb|smooth| to be elliptical. We selected all those with \verb|featured| greater than 0.3 as discs. \citetalias{2022MNRAS.509.3966W} used appearance-based questions rather than those requiring some physical interpretation. Hence, there is not an exact match between featured and disc galaxies, and similarly between smooth and elliptical. We further divided the disc sample by separating those with \verb|edge-on_yes| and \verb|edge-on_no| greater than 0.5 into edge-on and face-on, respectively. Finally, we separated those face-on into spiral and miscellaneous discs by taking a cut in the \verb|spiral| prediction at 0.5. These miscellaneous discs will include S0 galaxies \citepalias{2022MNRAS.509.3966W}, as well as potentially ring or irregular galaxies. While these cuts are rather crude, such as the lack of specific inclination to distinguish between face- and edge-on, they suffice for the broad categorisation required for this study. 

Our decision to loosen the restriction on the merger columns was only in part motivated by the need to increase the size of the control pool. We discovered that in the original stricter scenario, there were significantly fewer spiral galaxies (39) compared to edge-on discs (2172), a stark difference that was unexpected. Upon investigation, it became clear that this was a direct consequence of the choice to have all three of \verb|merging_minor|, \verb|merging_major|, and \verb|merging_merger| less than 0.08. Indeed, when this was increased to larger values, the proportion of spirals in the mix increased, as well as the total number of galaxies. It is unclear exactly why the classifier rated spirals as slightly more likely to be disturbed than edge-on galaxies, but it is possible that the spiral arms were mistaken for tidal disturbances. We evaluated the Pearson correlation coefficient and Spearman rank-order correlation coefficient on the \verb|spiral| prediction and combined \verb|merging| predictions (i.e. $1-$ \verb|merging_none|) from the \citetalias{2022MNRAS.509.3966W} catalogue. Indeed, we found a weak linear correlation and a moderate monotonic correlation between the spiral arm and disturbance predictions, with a significance level greater than 99 per cent.

\begin{table*}
    \caption{The number of galaxies in each morphological category. The left columns provide the breakdown of galaxies in each category from the \citetalias{2024MNRAS.534.1459G} DECaLS sample that were successfully cross-matched to the MPA-JHU SDSS data. The middle columns provide the same for the \citetalias{2013ApJ...765...28A} CFHTLS sample. The total pool column provides the number of galaxies available in the sample before matching the distributions of stellar mass and redshift between the tidal galaxies and controls. The last columns provide the total number of galaxies that passed the matching and selection phase. As the classes were non-exclusive, some galaxies with tidal features may have been counted twice or more; the number of these in each column is provided in brackets. Square brackets indicate the number of galaxies that were indicated to have both stream and linear or both shell and fan features in \citetalias{2013ApJ...765...28A}, and therefore only counted towards the stream or shell totals once.}
    \label{tab:numbers_of_galaxies}
    \begin{tabular}{c l c c l c c c c c c}
        \hline
        & \multicolumn{2}{c}{DECaLS} & & \multicolumn{2}{c}{CFHTLS} & & Total & \multicolumn{3}{c}{Selections} \\
        & Morphology & Count & & Morphology & Count & & pool & $M_\star$ \& $z$& PCA & BPT \\
        \hline
        \multirow{8}{*}{\rotatebox[origin=c]{90}{Tidal}} & \textit{Arm} & 252 & & Arm & 26 & & 278 & 271 & 219 & 204 \\
        & \multirow{2}{*}{\textit{Stream}} & \multirow{2}{*}{213} & \multirow{2}{*}{\huge{\{}} & Stream & 14 & \multirow{2}{*}{\huge{\}}} & \multirow{2}{*}{239} & \multirow{2}{*}{229} & \multirow{2}{*}{206} & \multirow{2}{*}{145} \\
        & & & & Linear & 12 [0]\\
        & \multirow{2}{*}{\textit{Shell}} & \multirow{2}{*}{46} & \multirow{2}{*}{\huge{\{}} & Shell & 6 & \multirow{2}{*}{\huge{\}}} & \multirow{2}{*}{61} & \multirow{2}{*}{56} & \multirow{2}{*}{55} & \multirow{2}{*}{28} \\
        & & & & Fan & 11 [2]\\
        & \textit{Diffuse} & 407 & & Miscellaneous & 11 & & 418 & 407 & 357 & 250 \\
        & Total & 760 & & Total & 63 & & 823 & 801 & 688 & 521 \\
        & (double, triple) & (156, 1) & & & (15, 1) & & (169, 2) & (158, 2) & (147, 1) & (104, 1) \\
        \hline
        \multirow{5}{*}{\rotatebox[origin=c]{90}{Non-tidal}} & Elliptical & 26117 & & Red sequence & 139 & & 26256 & 2658 & 2171 & 534 \\
        & Spiral & 5634 & & - & - &  & 5634 & 898 & 558 & 437 \\
        & Edge-on disc & 12382 & & - & - &  & 12382 & 1510 & 910 & 427 \\
        & Miscellaneous disc & 4122 & & Blue cloud & 192 &  & 4314 & 541 & 434 & 183 \\
        & Total & 48255 & & Total & 331 & & 48586 & 5607 & 4073 & 1581 \\
        \hline
        \multicolumn{2}{l}{Total} & 49015 & &  & 394 & & 49409 & 6408 & 4761 & 2102 \\
        \hline
    \end{tabular}
\end{table*}

The left-hand column of Table \ref{tab:numbers_of_galaxies} provides the number of DECaLS galaxies identified in each morphological category (after cross-matching to SDSS data, see Section \ref{sec:sdss}). Approximately 20 per cent (157) of the tidal galaxies had two or more different kinds of tidal features; the number of these is indicated in brackets in Table \ref{tab:numbers_of_galaxies}. For this work, we treated the classes as independent. Since some instances of a given class were also members of another, the total number of tidal galaxies was less than the sum of the number of galaxies with each feature. While the tidal galaxy sample could in principle also be divided based on bulk morphology, we lack sufficient statistics in this study to make this approach meaningful.

\subsection{CFHTLS sample}\label{sec:cfhtls}
To increase the size of the sample, we augmented the DECaLS sample by including galaxies with tidal features identified by \citet[hereinafter \citetalias{2013ApJ...765...28A}]{2013ApJ...765...28A}. To construct their sample, \citetalias{2013ApJ...765...28A} used images from the wide component of the Canada-France-Hawaii Telescope Legacy Survey \citep[CFHTLS;][]{2012AJ....143...38G}. The CFHTLS-wide survey used the MegaCam camera to image 150 deg$^2$ in five filters: \textit{u}$^*$\textit{g}$^\prime$\textit{r}$^\prime$\textit{i}$^\prime$\textit{z}$^\prime$. We estimated the \textit{r}$^\prime$-band limiting surface brightness (3$\sigma$ in 10$\times$10 arcsec$^2$ boxes) in the stacked data to be $\sim$29.4 mag arcsec$^{-2}$. \citetalias{2013ApJ...765...28A} visually inspected 1781 galaxies with magnitudes in the range $r^{\prime} \in (15.5, 17)$ and redshifts between $z \in (0.04, 0.2)$. They rated each galaxy with a confidence from 0 to 4 based on how likely it was to have a tidal feature. Each of the $\sim$300 galaxies that they indicated as likely to have a tidal feature, corresponding to confidence levels 3 and 4, was labelled using six different categories of tidal features: arms, fans, linear features, miscellaneous diffuse, shells, and streams. We took these labels and transformed the classes to those used in \citetalias{2024MNRAS.534.1459G}. In particular, we combined the fan and shell classes into the \textit{shell} class, and the stream and linear into just \textit{stream}. For the galaxies without tidal features, which we took as those with confidences 0 or 1 ($\sim$1300), \citetalias{2013ApJ...765...28A} assigned each to either the red sequence or blue cloud based on a cut in the ($g^\prime-r^\prime$) versus $M_{r^\prime}$ colour magnitude diagram. We assigned the red sequence galaxies to our elliptical category and the blue cloud to miscellaneous discs, within the control pool. To be consistent with the DECaLS sample and \citetalias{2024MNRAS.534.1459G}, we limited the absolute magnitudes of the CFHTLS galaxies to be between $-19\geq M_{r^\prime}\geq -22$. The middle column of Table \ref{tab:numbers_of_galaxies} indicates the number of each category identified in the CFHTLS (again after cross-matching to SDSS data) with their original label and how this corresponds to the DECaLS labels. Again, galaxies with multiple features are over-counted. The square brackets indicate galaxies with both shell and fan, or linear and stream features, which are counted only once towards the total.

\subsection{SDSS-derived data}\label{sec:sdss}
We cross-matched all of the galaxies across the combined DECaLS and CFHTLS samples to the SDSS-derived MPA-JHU DR7 catalogue\footnote{\url{https://wwwmpa.mpa-garching.mpg.de/SDSS/DR7/}} \citep{2003MNRAS.341...33K, 2004MNRAS.351.1151B} to provide estimates of stellar mass\footnote{Median total stellar mass.} and emission line fluxes\footnote{Provided in the MPA-JHU catalogue as the flux from a Gaussian fit to continuum-subtracted data.}. Every galaxy in the DECaLS sample had a corresponding match in the NSA (itself an SDSS-derived source), which included spectroscopic identifiers for SDSS spectra. We used these identifiers to cross-match the sample to the MPA-JHU. For the CFHTLS sample, we queried the SDSS DR7 photometric and spectroscopic tables to obtain matches within 3 arcsec and used the corresponding spectroscopic identifiers from the table to match to the MPA-JHU catalogue. For redshifts, we take the NSA-provided value for the DECaLS sample -- which were negligibly different from the MPA-JHU value -- and the MPA-JHU values for the CFHTLS galaxies as those provided by \citetalias{2013ApJ...765...28A} were photometric estimates.

Where there were duplicate entries for the same spectroscopic identifiers, we removed all of the duplicates for the DECaLS sample and kept only the closest match for CFHTLS. This typically corresponded to cases where two galaxies had small projected separations. We further cleaned the data by removing those galaxies where the fibre centre was more than 5 arcsec from the galaxy centre ($\sim$1200). Additionally, galaxies with no mass estimate in the MPA-JHU data were also removed ($\sim$3200).

\begin{figure}
    \includegraphics[width=\columnwidth]{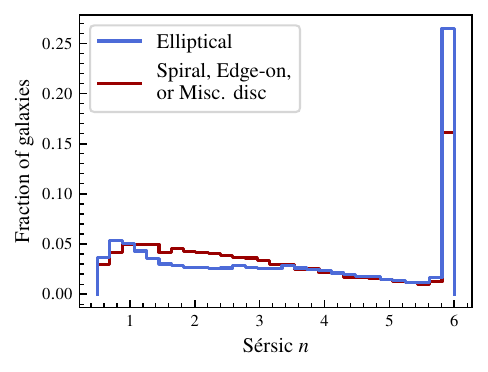}
    \caption{The normalised distribution of S{\'{e}}rsic indices for the DECaLS sample split between those identified as ellipticals and discs (spiral, edge-on, or miscellaneous) by the \citetalias{2022MNRAS.509.3966W} classifier.}
    \label{fig:sersic}
\end{figure}

The NSA provided estimates of the S{\'{e}}rsic indices for each of the galaxies in the DECaLS sample. Upon investigation, we found that some of the elliptical galaxies identified by the \citetalias{2022MNRAS.509.3966W} classifier had S{\'{e}}rsic indices more consistent with being discs (spiral, edge-on, or miscellaneous) and vice versa. Fig. \ref{fig:sersic} shows the distribution of S{\'{e}}rsic indices split between the elliptical and discs from the \citetalias{2022MNRAS.509.3966W} predictions. Either the original volunteers or the automated classifier could have misclassified these. However, this does serve as a reminder of the fallibility of machine learning algorithms; even a good algorithm will have some level of contamination in its predictions that should be accounted for. We inspected examples of both the ellipticals with low S{\'{e}}rsic indices, and discs with high ones, and decided to exclude the ellipticals that had indices less than 2.5 as contaminants. We retained all the discs in the control pool as those with high S{\'{e}}rsic indices appeared to simply be discs with prominent bulges. In total, this process removed $\sim$20$\,$000 galaxies and left $\sim$49$\,$000 galaxies available across both the tidal galaxies and control pool. The total pool column in Table \ref{tab:numbers_of_galaxies} provides the number of galaxies in each morphological category that were available and is the summation of both the DECaLS and CFHTLS columns.

\section{Selection criteria}\label{sec:selection}
\subsection{Stellar mass and redshift matching}\label{sec:massredshift}
\begin{figure}
    \includegraphics[width=\columnwidth]{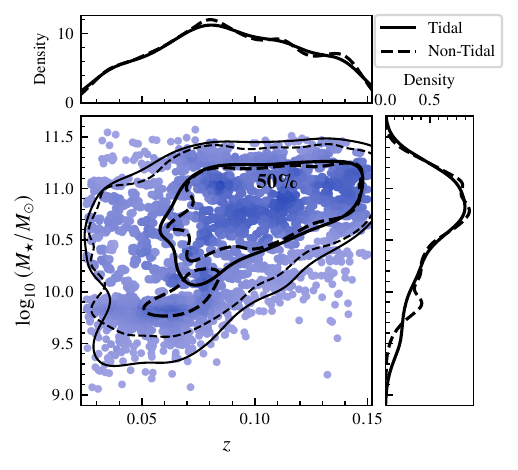}
    \caption{Stellar mass ($M_\star$)-redshift ($z$) distribution of the sample. Points are shaded such that darker points indicate a higher concentration of galaxies. The contours show the region containing 50 (thicker) and 90 (thinner) per cent of the data. The top and side panels show the individual kernel density estimates of stellar mass and redshift. The dashed and dotted lines indicate the distribution for the tidal sample and the non-tidal controls, respectively.}
    \label{fig:redshift_mass_dist}
\end{figure}

We imposed limits on the range of stellar masses and redshifts of the tidal galaxies and the control pool to be within $9.0 \leq \log_{10}\left(M_\star/M_\odot\right) \leq 11.6$ and $0.025 \leq z \leq 0.15$.  After this, we used a mass-redshift matching algorithm to ensure that the tidal galaxies and controls were selected with the same distributions in mass and redshift space. To build a sample of controls, we follow the process from \citetalias{2022MNRAS.516.4354W} and \citetalias{2022MNRAS.517L..92E}. First, for every tidal galaxy we selected the closest galaxy in $\log(M_\star/M_\odot)$ and $\log(z)$ space from the control pool, removing it from the pool. After this, we assessed the similarity between the mass and redshift distributions of the tidal and selected control samples using the two-sample Kolmogorov-Smirnov test. If the distributions were statistically similar with a probability greater than 5 per cent, we selected another round of controls from the pool, again removing those that were chosen from the pool. This process was repeated until the distributions became distinct, at which point we stopped and restored the controls to the previous step, where the samples were similar. We then verified that each tidal galaxy had at least one matching control within $|\Delta\log(M_\star/M_\odot)|\leq0.2$ and $|\Delta z|\leq0.02$. Overall, the final sample consisted of $\sim$6400 galaxies, with seven control galaxies (hereinafter non-tidal galaxies) for each tidal galaxy. The $M_\star$ \& $z$ column in Table \ref{tab:numbers_of_galaxies} provides the total number of galaxies that were selected by this process across the different morphologies. Fig. \ref{fig:redshift_mass_dist} provides the stellar mass-redshift distribution of the whole sample with points shaded such that darker points indicate a higher concentration of galaxies. The figure also shows the distributions split between the tidal galaxies and the non-tidal controls.

\subsection{Evolutionary phase}\label{sec:evophaseselect}
\begin{figure}
    \includegraphics[width=\columnwidth]{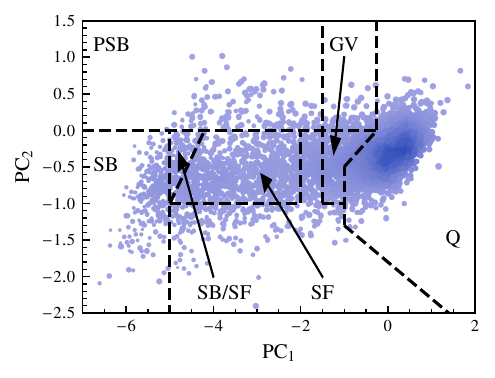}
    \caption{The evolutionary phase based on the \citetalias{2007MNRAS.381..543W} principal component analysis (PCA). See the text for details of the selection criteria based on the first and second principal components (PC$_1$ and PC$_2$). The galaxies are separated into those identified as post-starbursts (PSB), starbursts (SB), star-forming (SF), those intermediate between SB and SF (SB/SF), those in the green valley (GV), and those that are quiescent (Q). Galaxies that were outside of these classifications were considered as not specified (NS). Where the error on PC$_1$ or PC$_2$ crossed the boundary of a phase, the galaxy was considered as unidentified (U).}
    \label{fig:psb_selection}
\end{figure}

To identify the evolutionary phase of the galaxy, we cross-matched the sample to the \citet[][hereinafter \citetalias{2007MNRAS.381..543W}]{2007MNRAS.381..543W} principal component analysis (PCA) catalogue\footnote{\url{http://star-www.st-andrews.ac.uk/~web2vw8/downloads/DR7PCA.html}}, again using spectroscopic identifiers to do so. The \citetalias{2007MNRAS.381..543W} sample consisted of $\sim$900$\,$000 galaxies with optical SDSS DR7 spectra. In the analysis, the spectrum of a galaxy was decomposed into a combination of different eigenspectra, which were determined by the training process. The amount each eigenspectrum contributed to the overall spectrum is the principal component. \citetalias{2007MNRAS.381..543W} indicated that the first principal component ($\text{PC}_1$) is a direct tracer of the D$_n4000$ index, itself tracing the age of the stellar population \citep{1983ApJ...273..105B, 1985ApJ...297..371H, 1999ApJ...527...54B}. Combining both $\text{PC}_1$ and the second principal component ($\text{PC}_2$) gives a measure of the H$\delta$ equivalent width \citepalias{2007MNRAS.381..543W}. Both the D$_n4000$ and H$\delta$ can be used to constrain the stellar age and the amount of recent bursty star formation in galaxies \citep{2003MNRAS.341...33K}, and thus $\text{PC}_1$ and $\text{PC}_2$ can also be used to determine these.

We followed previous works that have used the PCA catalogue to study post-starbursts \citepalias{2018MNRAS.477.1708P, 2022MNRAS.516.4354W, 2022MNRAS.517L..92E} in our selection of post-starburst and star-forming galaxies. We limited the spectral signal-to-noise ratio (SNR) in the $g^{\prime}$-band to be greater than 8, to ensure adequate representation in the PCA space, and then applied cuts to the first two principal components ($\text{PC}_1$ and $\text{PC}_2$) and their errors ($\Delta\text{PC}_1$ and $\Delta\text{PC}_2$) such that those with
\begin{equation}
\begin{aligned}
    \text{PC}_1 + \Delta\text{PC}_1 &< -1.5\\
    \text{PC}_2 - \Delta\text{PC}_2 &> 0.0
\end{aligned}
\end{equation}
were identified as post-starbursts (PSB). Similarly, galaxies in the region defined by
\begin{equation}
\begin{aligned}
    \text{PC}_1 + \Delta \text{PC}_1 &< -2.0\\
    \text{PC}_1 - \Delta \text{PC}_1 &> -5.0\\
    \text{PC}_2 + \Delta \text{PC}_2 &< 0.0\\
    \text{PC}_2 - \Delta \text{PC}_2 &> -1.0
\end{aligned}
\end{equation}
were taken to be star-forming (SF).

For the remaining categories, we follow the regions indicated by \citetalias{2007MNRAS.381..543W} and select quiescent galaxies (Q), starbursts (SB), and those in the green valley (GV). We note that there was a region of the parameter space where \citetalias{2022MNRAS.516.4354W} and \citetalias{2022MNRAS.517L..92E} indicated that the galaxies would be star-forming, but \citetalias{2007MNRAS.381..543W} indicated as starbursts. We assign galaxies in this region as an intermediate class (SB/SF) to resolve this issue. Finally, we defined two other categories based on the PCA criteria. First, we assigned all galaxies that were not within the boundary of a class to the not specified (NS) class. Secondly, where the error bars on PC$_1$ or PC$_2$ crossed a boundary line, we assigned that galaxy to the unidentified (U) category. This followed both \citetalias{2022MNRAS.516.4354W} and \citetalias{2022MNRAS.517L..92E}, who eliminated galaxies where the error bars crossed the boundary between classes, hence the requirements above involving the error on the components. Figure \ref{fig:psb_selection} shows how these phases were selected based on the above criteria and the principal components.

Again following \citetalias{2018MNRAS.477.1708P}, \citetalias{2022MNRAS.516.4354W} and \citetalias{2022MNRAS.517L..92E}, we removed PSBs with a high dust content. As noted in \citetalias{2022MNRAS.516.4354W}, it is unclear whether these dusty systems are genuine post-starbursts or if optical spectral signatures of ongoing star formation, such as H$\alpha$ or [\ion{O}{ii}] emission, are suppressed by the significant amount of dust \citep[see, e.g.][]{1999ApJ...525..609S, 2000ApJ...529..157P, 2001ApJ...554L..25M, 2004A&A...427..125G}. We employed the Balmer decrement selection from \citetalias{2018MNRAS.477.1708P} and \citetalias{2022MNRAS.516.4354W} to remove these dusty systems. Those galaxies with Balmer signal-to-noise ratios
\begin{equation}
    \text{SNR}_{\text{balmer}} = \frac{1}{\sqrt{\frac{1}{\text{SNR}_{\text{H}\alpha}^{2}} + \frac{1}{\text{SNR}_{\text{H}\beta}^{2}}}} > 3
\end{equation}
and a Balmer decrement
\begin{equation}
    D_{\text{balmer}} = \frac{\text{H}\alpha}{\text{H}\beta} > \begin{cases}
        6.6, \; \text{if} \; M_\star > 3\times10^{10} M_\odot\\
        5.2, \; \text{if} \; M_\star < 3\times10^{10} M_\odot
    \end{cases}
\end{equation}
were removed from the sample. In total, 4761 galaxies passed all the PCA selection criteria, with only 55 having been removed as dusty systems. The PCA column in Table \ref{tab:numbers_of_galaxies} provides the breakdown of these galaxies into their different morphologies.

\subsection{Active galactic nuclei}\label{sec:agnselect}
\begin{figure}
\includegraphics[width=\columnwidth]{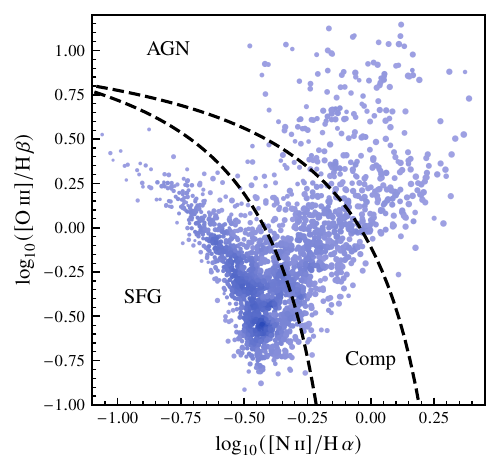}
    \caption{Baldwin–Phillips–Terlevich \citep[BPT;][]{1981PASP...93....5B} optical emission line analysis for the determination of active galactic nuclei (AGN) activity. The sample was split into galaxies with AGN, star-forming galaxies (SFG), and composites (Comp) using the \citet{2006MNRAS.372..961K} criteria.}
    \label{fig:agn_selection}
\end{figure}
We also determined which galaxies showed evidence of AGN activity using the Baldwin–Phillips–Terlevich \citep[BPT;][]{1981PASP...93....5B} optical emission line analysis with the H$\alpha$, H$\beta$, [\ion{O}{iii}] $\lambda$5007{\AA}, and [\ion{N}{ii}] $\lambda$6584{\AA} lines. Taking the line fluxes and errors from the MPA-JHU catalogue, we imposed a minimum SNR of 3 for each of the four lines. We chose this over a higher SNR threshold to balance the completeness and contamination within the AGN sample, with a preference to be more complete, whilst maintaining a sufficient degree of purity. We note that, as a consequence, our AGN sample will include a greater contribution of low-ionisation nuclear emission-line regions \citep[LINERs; see, e.g.][who remove a high proportion of these with an SNR cut]{2019MNRAS.487.2491E} than if we had applied a higher threshold. There is not necessarily a clear consensus within the literature whether these LINERs correspond to AGN, with some studies saying they do not \citep[e.g.][]{2012ApJ...747...61Y, 2016MNRAS.461.3111B, 2025ApJ...984..106N} and others including them in their AGN sample \citep[e.g.][]{2012AJ....144...11M, 2015MNRAS.447..110S}. We elected to include these LINERs in our AGN sample, although we note that this may introduce contamination from emission not associated with an AGN. From this sample, we used the \citet{2006MNRAS.372..961K} criteria to identify which galaxies hosted AGN, which were star-forming, and which were composites. Specifically, galaxies with AGN activity had an [\ion{O}{iii}] to H$\beta$ flux ratio above 
\begin{equation}
    \log_{10}\left([\ion{O}{iii}]/\text{H}\beta\right) > \frac{0.61}{\log_{10}\left([\ion{N}{ii}]/\text{H}\alpha\right)-0.47} + 1.19\\
\end{equation}
and those with a ratio below
\begin{equation}
    \log_{10}\left([\ion{O}{iii}]/\text{H}\beta\right) < \frac{0.61}{\log_{10}\left([\ion{N}{ii}]/\text{H}\alpha\right)-0.05} + 1.3
\end{equation}
were considered as star-forming (we denote this as SFG to avoid confusion with SF from the PCA selection). According to the \citet{2006MNRAS.372..961K} scheme, everything in between these is considered a composite (Comp) between AGN and SFG. Given that Comp is a hybrid between AGN and SFG, we elected not to remove galaxies where the error bars crossed the classification boundaries, as we did with the evolutionary phases. Fig. \ref{fig:agn_selection} shows how the different categories were determined from the BPT diagram for our sample of galaxies. In all, 2102 galaxies passed the selection criteria, and this can be seen in Table \ref{tab:numbers_of_galaxies}.

\section{Results}\label{sec:results}

\subsection{Evolutionary phase}\label{sec:evophase}
\begin{figure*}
    \includegraphics[width=\textwidth]{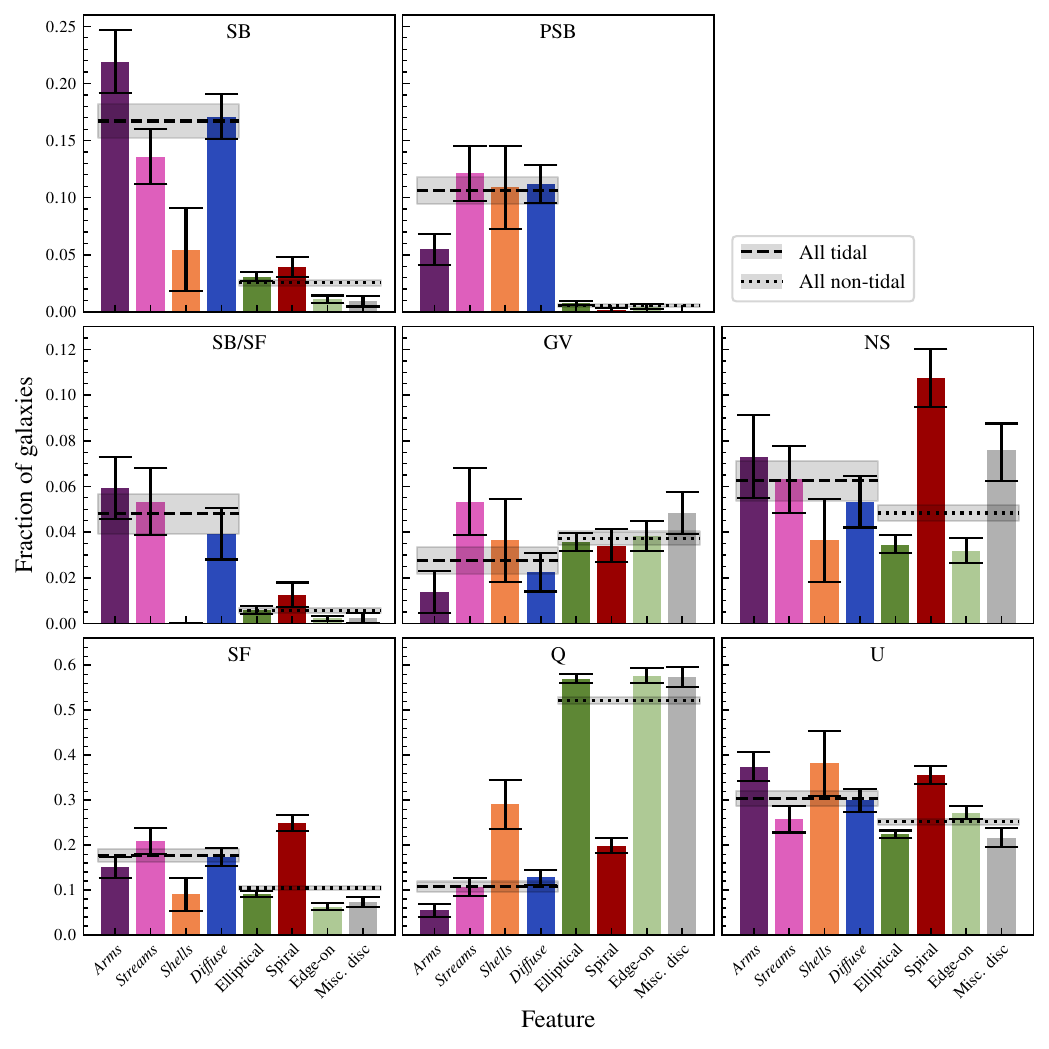}
    \caption{The fraction of each galaxy category in the phases identified by the \citetalias{2007MNRAS.381..543W} PCA analysis. The phases were post-starbursts (PSB), starbursts (SB), star-forming (SF), intermediate star-forming and starburst (SB/SF), the green valley (GV), quiescent (Q), not specified (NS), and unidentified (U); see the text for details. Each bar shows the median fraction of that kind of galaxy identified in that phase. Thus, the normalisation is such that the total in each feature should be one across all the panels. The dashed line indicates the fraction summed over all tidal classes, and similarly, the dotted line provides the fraction over all non-tidal classes. The medians and errors were determined using the bootstrap method (see text).}
    \label{fig:psb_fraction}
\end{figure*}

Fig. \ref{fig:psb_fraction} shows the fraction of galaxies that were identified to be in each evolutionary phase (panels) as a function of morphological class ($x$-axis). For example, the top left panel indicates that approximately 22$\pm$3 per cent of the galaxies with \textit{arm} features were SB, while the top middle panel shows that around 11$\pm$2 per cent of the \textit{diffuse} galaxies were PSB. The dashed lines provide the fraction of tidal galaxies, summed over all feature classes, in the different phases, and the dotted line shows the same thing for the non-tidal galaxies. Each of the calculated fractions is a median estimated using the bootstrap method \citep{1979AoS...E} along with the corresponding error. The method worked as follows. For a given morphological class, $N_{c}$ galaxies were randomly selected with replacement, where $N_{c}$ was the total number of galaxies in that class (see Table \ref{tab:pca_numbers}). We then determined the fraction of galaxies in each evolutionary phase for that particular sampling. The random selection was repeated $100\,000$ times to build a distribution of fractions, at which point the median and 68.5 per cent confidence interval were determined. Table \ref{tab:pca_numbers} provides the actual numbers of galaxies determined to be in each evolutionary phase across all morphologies, including the total number of tidal and non-tidal galaxies.

Several interesting results are identifiable in Fig. \ref{fig:psb_fraction}. Firstly, there is a striking difference between the tidal and non-tidal galaxies identified in both the SB and PSB phases, with each category of tidal feature having a significantly larger fraction of galaxies in these phases. Overall, tidal galaxies were 19.6$\pm$5.0 times more likely to be in a PSB phase than their non-tidal controls. For SB, this decreased to 6.6$\pm$0.9 times more likely. Secondly, the fractions vary depending on the type of tidal feature in the galaxy. For example, galaxies with \textit{shell} features were more likely to be Q than the other tidal galaxies and, similarly, were less likely to be SB. On the other hand, \textit{Arm} features were more likely to be in an SB phase and less likely to be PSB.

\begin{table*}
    \caption{The number of galaxies identified in each of the PCA categories.}
    \label{tab:pca_numbers}
    \begin{tabular}{c l c c c c c c c c c}
        \hline
        \multicolumn{2}{l}{\multirow{2}{*}{Morphology}} & \multicolumn{8}{c}{PCA category} & \multirow{2}{*}{Total} \\
        & & PSB & SB & SB/SF & SF & GV & Q & NS & U &\\
        \hline
        \multirow{5}{*}{\rotatebox[origin=c]{90}{Tidal}} & \textit{Arms} & 12 & 48 & 13 & 33 & 3 & 12 & 16 & 82 & 219\\
        & \textit{Streams} & 25 & 28 & 11 & 43 & 11 & 22 & 13 & 53 & 206\\
        & \textit{Shells} & 6 & 3 & 0 & 5 & 2 & 16 & 2 & 21 & 55\\
        & \textit{Diffuse} & 40 & 61 & 14 & 62 & 8 & 46 & 19 & 107 & 357\\
        & Total & 73 & 115 & 33 & 122 & 19 & 74 & 43 & 209 & 688\\
        \hline
        \multirow{5}{*}{\rotatebox[origin=c]{90}{Non-tidal}} & Elliptical & 17 & 67 & 13 & 198 & 77 & 1238 & 75 & 486 & 2171\\
        & Spiral & 1 & 22 & 7 & 139 & 19 & 111 & 60 & 199 & 558\\
        & Edge-on disc & 4 & 10 & 2 & 57 & 35 & 525 & 29 & 248 & 910\\
        & Miscellaneous disc & 0 & 4 & 1 & 32 & 21 & 249 & 33 & 94 & 434\\
        & Total & 22 & 103 & 23 & 426 & 152 & 2123 & 197 & 1027 & 4073\\
        \hline
        \multicolumn{2}{l}{Total} & 95 & 218 & 56 & 548 & 171 & 2197 & 240 & 1236 & 4761 \\
        \hline
    \end{tabular}
\end{table*}

Fig. \ref{fig:psb_fraction} also recovers many expected results, lending credibility to our analysis. In particular, a large proportion of spiral galaxies ($\sim$30 per cent) were SF, SB or SB/SF, and the majority of ellipticals were Q galaxies (57.0$\pm$1.1 per cent). The fraction indicated as U was broadly constant across all classes, but was slightly higher in the tidal galaxies. This was also largely true for those identified as NS, although we note that some of the disc-like galaxies (spiral and miscellaneous disc) showed greater variance than the other classes. This may be due to the inclusion of all of the blue \citetalias{2013ApJ...765...28A} galaxies into the miscellaneous disc class. There is, however, a somewhat surprising result: a high proportion of disc galaxies were Q. This result is explored more in Section \ref{sec:qui_discs}.

To ensure that our results were not driven by any one sample, we applied our analysis independently to the DECaLS and CFHTLS galaxies. Reassuringly, we found that both samples obtained the same result: the tidal galaxies exhibited an excess of starburst and post-starburst galaxies compared to the non-tidal controls.

\subsection{AGN fraction}\label{sec:bptresults}
\begin{figure*}
    \includegraphics[width=\textwidth]{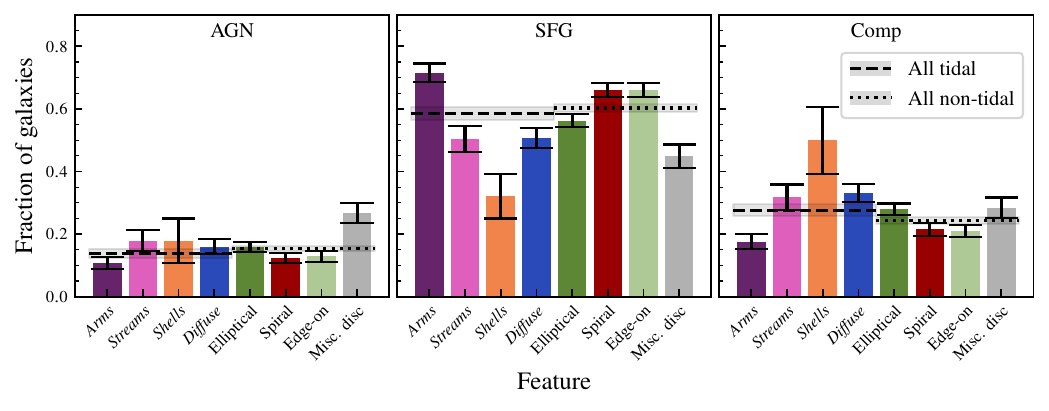}
    \caption{The fraction of each category of galaxy in each of the BPT classes. Each bar shows the median fraction of that morphology that was identified to host AGN (left), be star-forming (centre), or a composite (Comp, right). The dashed line indicates the all tidal fraction and, similarly, the dotted all non-tidal. The medians and errors were determined using the bootstrap method (see text).}
    \label{fig:agn_fraction}
\end{figure*}

In a similar manner, Fig. \ref{fig:agn_fraction} provides the fraction of each type of galaxy identified in each BPT class. In contrast to the PCA phases, there appeared to be no statistically significant difference in the AGN fraction for tidal versus non-tidal galaxies. The difference between the tidal and non-tidal AGN fractions was only 0.016$\pm$0.018. Again, the errors in Fig. \ref{fig:agn_fraction} were determined using the bootstrap method, with the distribution determined over $100\,000$ samples. Table \ref{tab:bpt_numbers} provides the total number of each morphology identified as hosting AGN activity or not.

There was some slight variation between the different tidal classes. \textit{Stream} and \textit{shell} features appear to have a marginally higher fraction of AGN compared to the other classes, whereas \textit{arms} seemed to have a somewhat lower fraction. However, these differences were within the uncertainties, and as such, the result was not significant. The same could be said for spiral and edge-on galaxies, which also showed a slightly reduced fraction of AGN, but again, this can simply be attributed to the uncertainties. 

\begin{table}
    \caption{The number of galaxies identified in each of the BPT categories.}
    \label{tab:bpt_numbers}
    \begin{tabular}{c l c c c c}
        \hline
        \multicolumn{2}{l}{\multirow{2}{*}{Morphology}} & \multicolumn{3}{c}{BPT category} & \multirow{2}{*}{Total}\\
        & & AGN & Comp & SFG \\
        \hline
        \multirow{5}{*}{\rotatebox[origin=c]{90}{Tidal}} & \textit{Arms} & 22 & 36 & 146 & 204\\
        & \textit{Streams} & 26 & 46 & 73 & 145\\
        & \textit{Shells} & 5 & 14 & 9 & 28\\
        & \textit{Diffuse} & 40 & 83 & 127 & 250\\
        & Total & 72 & 144 & 305 & 521\\
        \hline
        \multirow{5}{*}{\rotatebox[origin=c]{90}{Non-tidal}} & Elliptical & 85 & 149 & 300 & 534\\
        & Spiral &  54 & 94 & 289 & 437\\
        & Edge-on disc & 55 & 90 & 282 & 427\\
        & Misc. disc & 49 & 52 & 82 & 183\\
        & Total & 243 & 385 & 953 & 1581\\
        \hline
        \multicolumn{2}{l}{Total} & 315 & 529 & 1258 & 2102 \\
        \hline
    \end{tabular}
\end{table}

\section{Discussion}\label{sec:discussion}

\subsection{Evolutionary Phase}
Our results indicate that the presence of faint tidal features is linked to intense periods of star formation and the subsequent quenching of this. This result is broadly consistent with previous works, which have considered tidal features irrespective of their particular morphologies (e.g. \citealt{2011MNRAS.412..591P}; \citetalias{2018MNRAS.477.1708P}).  Of particular relevance is the study of \citetalias{2022MNRAS.517L..92E}, on which we have largely based our approach. They identified post-starbursts in galaxies with stellar masses $\log_{10}\left(M_\star/M_\odot\right) > 10.0$ and post-merger features from the Canada France Imaging Survey \citep[CFIS;][]{2017ApJ...848..128I}. As mentioned, \citetalias{2022MNRAS.517L..92E} also used the \citetalias{2007MNRAS.381..543W} PCA catalogue to determine which galaxies were post-starburst. They found that 20$\pm$2 per cent of their post-merger galaxies were post-starbursts, which was roughly 30 times greater than their non-merging controls. We report a lower fraction of 10.6$\pm$1.2 per cent (across all the tidal galaxies), approximately 20 times greater than that of the controls. We note that while the vast majority of our galaxies did not have any obvious companion, some did, whereas \citetalias{2022MNRAS.517L..92E} required that each of their galaxies was post-coalescence. Furthermore, they required that the mass ratio of the merger was at least 10:1, while we imposed no such selection. In addition, while our sample is slightly larger than \citetalias{2022MNRAS.517L..92E} (688 vs 508),  our limiting surface brightness is slightly lower, on average, than that of the CFIS dataset they used ($\sim 28$ mag arcsec$^{-2}$ vs $\sim 28.7$ mag arcsec$^{-2}$, see \citealt{2025MNRAS.tmp.1100S}). Any one of these factors, or others, may have influenced the slightly different enhancement factors found. 

Our results for the starbursts are broadly consistent with those in the literature that find an enhancement to the star formation rate in post-merger galaxies \citep[see, e.g.][]{2022MNRAS.514.3294B, 2025MNRAS.538L..31F}. \citet{2013MNRAS.435.3627E} investigated the fraction of post-merger galaxies that showed evidence of ongoing star formation, and starbursts, in SDSS. Of the 97 galaxies with post-merger features that they found in the \citet{2010MNRAS.401.1552D} catalogue, $\sim$40 per cent were actively star forming, with the post-merger sample being roughly 10 times more likely to be in a starburst phase compared to controls. \citet{2018ApJ...868...46S} found that around 20 per cent of the merging galaxies in CANDELS/3D-HST were in a starburst phase. Although \citet{2018ApJ...868...46S} considered a wider redshift range and pre-coalescence systems, our results are comparable with both their work and \citet{2013MNRAS.435.3627E}. Overall, we too recover that $\sim$40 per cent of the tidal sample is actively forming stars in some manner (SB, SF, and SB/SF) and obtain a similar fraction of starbursts ($\sim$20 per cent for SB and SB/SF).

Compared to previous studies, a novel aspect of our work is the sensitivity to different tidal feature morphologies. Our results provide some of the first empirical evidence that the star formation behaviours may have a dependence on the detailed nature of the interaction, as traced by the tidal feature class. For example, \textit{shell} galaxies showed lower levels of star formation activity than the other tidal categories, albeit still enhanced compared to the non-tidal controls. In particular, they were 2.3$\pm$0.5 times more likely to have a quiescent classification Q than \textit{diffuse} galaxies and 5.3$\pm$1.7 times more likely than those with \textit{arms}. $N$-body and hydrodynamical simulations of mergers that produce shells suggest that the features only become visible after the star formation has quenched and that they remain visible for longer than other types of features \citep[e.g.][]{2018MNRAS4801715P,2023MNRAS.518.3261P}. Our results are broadly consistent with this, supporting the idea that merger-driven star formation in shell galaxies generally shuts down more rapidly than the timescale on which the shells lose their spatial coherence. The \textit{shell} systems that we identified as presently forming stars may be systems that are caught very shortly after the merger event. With a sufficiently large sample of shell galaxies and their associated star formation indicators, it might be possible to directly test this idea through correlating the prominence of the shell system (e.g. surface brightness, mass) with the evolutionary phase. 

Conversely, galaxies with tidal \textit{streams} were 1.7$\pm$0.3 times more likely to be actively star forming than not (i.e. SF, SB, and SB/SF vs Q and PSB). This was also the case for galaxies with \textit{arms}, to an even greater degree, at 3.9$\pm$0.8 more likely, with SB being the most common phase. Indeed, \textit{arm} galaxies were 1.3$\pm$0.2 times more likely to be SB than the tidal population overall, and 4.0$\pm$2.7 times more likely than the \textit{shell} population. This difference between galaxies with \textit{streams} and \textit{arms} may arise from the nature of these events.  In the \citetalias{2024MNRAS.534.1459G} classification scheme,  \textit{arms} and \textit{streams} have similar morphologies, but their origins are expected to be rather distinct. In particular, \textit{arms} are fairly bright features that are expected to arise as a result of major mergers \citep[e.g.][]{1972ApJ...178..623T, 1992ApJ393484B} or close fly-by interactions with another massive galaxy \citep[e.g.][]{2012ApJ...751...17S}. These are significant events that will cause large perturbations to the host galaxy's potential. They will likely induce strong gas dynamical effects, leading to gas inflow and enhanced central star formation \citep[e.g.][]{1994ApJ...425L..13M, 1996ApJ...464..641M, 2014MNRAS.442L..33R}. On the other hand, the features that we classify as \textit{streams} are faint and narrow, consistent with their being produced by the tidal disruption of dwarf satellites as they orbit a massive host. These represent minor or mini-mergers, and while they can still impact the star formation properties of the host galaxy, their effect is likely to be smaller \citep[e.g.][]{2020NatAs...4..965R}. 

Star-forming galaxies are known to exhibit a correlation between their star formation rates and their stellar masses \citep[e.g.][]{2004MNRAS.351.1151B,2007ApJS..173..267S}. It is therefore important to investigate if this underlying relationship could be the cause of any of the trends we see in the star formation properties of different classes of tidal galaxies. As discussed in Section \ref{sec:massredshift}, the tidal galaxy sample was selected to lie in the mass range spanning $9.0 \leq \log_{10}\left(M_\star/M_\odot\right) \leq 11.6$, and the redshift range spanning $0.025 \leq z \leq 0.15$. Within this mass range, we examined whether there were statistically significant differences in the mass distributions of galaxies belonging to different tidal classes. We found that \textit{stream}, \textit{shell}, and \textit{diffuse} galaxies were drawn from similar distributions, but the masses for the \textit{arm} galaxies were statistically smaller. However, even though the \textit{arm} galaxies were less massive, we found them to be more likely to be star-forming or bursting, contrary to the established trend where lower-mass galaxies have lower star formation rates. We therefore conclude that the trends we observed between tidal feature morphology and the levels of ongoing and recent star formation genuinely reflect the effects of the merger events from which the features originated.

\subsection{AGN fraction}\label{sec:agndiscuss}
In this work, we presented a brief look at the AGN fraction in our sample of tidal galaxies. Indeed, much debate still exists regarding whether or not mergers play a role in fueling AGN activity \citep[see, e.g.][]{2023OJAp....6E..34V, 2025OJAp....8E..12E}. Recent works have studied the AGN fraction in post-merger galaxies in the nearby Universe in a similar manner to us, although again without sensitivity to tidal feature morphology. \citet{2023MNRAS.519.6149B} looked at optical emission line selected AGN in $z < 0.5$ galaxies identified in CFIS imaging. They selected their post-merger sample using a combination of machine learning and visual inspection, and also used MPA-JHU emission line fluxes. Overall, they found an excess frequency of AGN in their sample by a factor of 3.7, when compared to mass and redshift-matched controls. \citet{2023ApJ...944..168L} studied the excess of optically selected AGN in high mass ($\log_{10}M_\star/M_\odot\geq10.5$) post-merger galaxies with X-ray detections from Chandra or XMM-Newton, again comparing to X-ray detected controls matched in mass and redshift. When including both Seyferts and LINERs they found an AGN excess factor of 1.8 in the post-merger galaxies. Similarly, \citet{2024MNRAS.533.3068B} used the same process as \citet{2023MNRAS.519.6149B} but instead identified post-merger galaxies in DECaLS data. Again, for their optically selected narrow-line AGN (using emission lines from the MPA-JHU) they found an excess, with AGN being around 4.5 times more frequent in the post-merger sample. 

In contrast, we found no overall excess of AGN in galaxies with tidal features, seemingly contradicting these studies and supporting studies that find no link. Furthermore, we saw no evidence that the AGN fraction had a dependence on the particular morphology of the feature. While we do note that \textit{arms} seemed to have a smaller fraction of AGN compared to the other tidal categories, this was not significant when considering the error bars. Taken at face value, this result could imply that the merger processes we are sensitive to in this study are not significant enough to drive the growth of the black hole, or that the timescale on which the AGN is switched on differs from the timescale over which the tidal features remain observable.

On the other hand, there are other reasons that could explain why we do not detect excess AGN activity in our sample relative to controls. Firstly, we have
adopted a method of AGN identification based on the strength of optical emission lines, but this may have missed obscured AGN \citep[e.g.][]{2008ApJ...677..926S}.  Indeed, when studying AGN selected using both optical SDSS spectra and photometric colours from the Wide-field Infrared Survey Explorer, \citet{2017MNRAS.464.3882W} found that merging galaxies are at least five times more likely to host obscured AGN than non-merging controls.  However, when considering solely the frequency of optically-selected AGN, they found no excess in mergers compared to non-mergers, as in the present work.  Exploring the AGN fraction in our sample of tidal galaxies using various other tracers of nuclear activity is beyond the scope of this paper, but is a potential future avenue. 

Other aspects that may be relevant are the mass range of our sample, our SNR cut, and the way in which we normalise our AGN detection rates. Considering the former, previous work has shown that the excess AGN fraction in galaxies with disturbances was less significant in galaxies with $\log_{10}\left(M_\star/M_\odot\right) \lesssim 10.7$ \citep{2019MNRAS.487.2491E}. As roughly half of the present sample lies below this value (see Fig.\ref{fig:redshift_mass_dist}), this may have contributed to diluting any signal. To test this, we recalculated the fractions using only our tidal and non-tidal galaxies that had masses $\log_{10}\left(M_\star/M_\odot\right) \geq 10.7$. The resulting 203 tidal galaxies had a roughly consistent distribution of morphologies to the original sample, and the 405 non-tidal galaxies had roughly equal contributions from each category. While we did find that the gap between fractions of tidal and non-tidal galaxies hosting AGN grew in this higher mass sample, the frequency of AGN was actually higher in the non-tidal galaxies. Therefore, we suggest that the mass limits do not have an impact on the lack of signal. Considering the impact of the emission-line SNR cut, we found that varying this value between 3 and 6 had a negligible impact on the recovered result, as we still found no excess in the tidal galaxies. This indicates that our result was not impacted by a greater contribution from LINERs, which were systematically removed to a greater extent than Seyferts when varying the SNR cut.

Finally, regarding our approach to the normalisation of detection rates, we note that some studies define an AGN fraction by dividing by the total number of galaxies in the sample regardless of whether they passed a SNR cut or not \citep[e.g.][]{2019MNRAS.487.2491E, 2023ApJ...944..168L, 2025A&A..L}, while we have instead normalised by exclusively using those that passed the BPT SNR cut (see Section \ref{sec:agnselect}). Our AGN frequency in tidal galaxies was, on the whole, 0.90$\pm$0.11 times that of the non-tidal controls. When considering instead the AGN frequency as a function of all galaxies, not just those that passed the SNR cut, this increased to 2.1$\pm$0.3. These results are explored further in Appendix \ref{app:altagn}. While this value is considerably higher and statistically significant, it is less secure since we cannot confidently rule out the presence of active galaxies with poor quality spectra in the sample that failed the SNR cut. Studies that include these galaxies often rely on the assumption that these galaxies are passive or non-star-forming. However, upon investigation, we find that this is not universally true. Approximately 6 per cent of the galaxies that failed the BPT SNR cut were successfully identified as star-forming in some manner by the PCA analysis, with $\sim$65 per cent being Q. Given this, we therefore choose not to rely on the assumption that the galaxies with low SNR in the emission lines are passive. We suspect, then, that this difference in normalisation may be a significant driver of the differences between our study and studies that find such a signal.

\subsection{Non-tidal galaxies}

In Section~\ref{sec:evophase}, we noted an unexpected result in our analysis of the correlation between galaxy morphology and evolutionary phase, namely the high fractions of quiescent discs. We briefly explore the origin of this result here, as well as the contrasting case of SF galaxies within the elliptical control sample. 

\subsubsection{Quiescent disc galaxies}\label{sec:qui_discs}
The high fraction of Q discs is largely driven by the edge-on ($\sim$57 per cent, see Fig.~\ref{fig:psb_fraction}) and miscellaneous disc (also $\sim$57 per cent) populations. For the edge-on discs, the Q designation could plausibly be a consequence of observing the galaxy through the disc with significant dust extinction \citep[see, e.g.][]{2010MNRAS.404..792M}. Additionally, for both the edge-on and miscellaneous discs, as well as the Q spirals ($\sim$20 per cent), aperture effects due to the size of the SDSS fibre are likely to be relevant. We investigated how much of the galaxy was covered by the 3-arcsecond SDSS fibre diameter \citep{2000AJ....120.1579Y} in the low redshift range of this study. We found that each aperture primarily covered the galaxies' central bulge; indeed, even at the maximum redshift limit ($z\sim0.15$), the aperture still only captured the central light ($\sim$0.8-4 kpc; assuming the Planck Collaboration VI \citeyear{2020A&A...641A...6P} cosmology). Since the stellar population of the bulge is generally thought to be older than the disc \citep[see, e.g.][]{2014A&A...570A...6S,2015A&A...581A.103G,2017MNRAS.466.4731G,2017MNRAS.465.4572Z}, although this is contested by some sources \citep[see, e.g.][for a summary]{2023PASA...40....2L}, this may explain why a significant proportion of disc systems were identified as Q. We further note that the way that \citetalias{2022MNRAS.509.3966W} identifies S0 galaxies, which are established to have lower SFRs than other discs \citep{2004A&A...414...23J}, would place them in our miscellaneous disc class.

\subsubsection{Star-forming elliptical galaxies}\label{sec:sf_ellips}
We found that a number of the elliptical galaxies (9.1$\pm$0.6 per cent) were in the SF phase (see Fig.~\ref{fig:psb_fraction} and Table~\ref{tab:pca_numbers}). This was slightly surprising; however, we expect that some ellipticals may truly be star-forming \citep[see, e.g.][]{2004ApJ...601L.127F, 2005ApJ...619L.111Y, 2023A&A...669A..11P}, and the fraction we recovered is broadly consistent with that expected from the literature \citep[5-8 per cent; see, e.g.][]{2009MNRAS.396..818S, 2016A&A...588A..79L, 2022MNRAS.509..550J}. We note that this fraction also had a dependence on our choice of S{\'{e}}rsic index. We observed a higher fraction of SF elliptical galaxies when we did not impose a restriction on the  S{\'{e}}rsic index, suggesting that there could be some residual contamination from discs driving this result.  

\section{Summary}\label{sec:summary}   
Faint tidal features are expected to be produced during mergers and interactions between galaxies. As well as leaving behind spectacular debris, these interactions are thought to drive inflows of gas to the nucleus \citep[e.g.][]{1996ApJ...471..115B}, which can then trigger intense star formation \citep[e.g.][]{2006MNRAS.373.1013C}. In this work, we have studied the level of star formation and AGN activity in galaxies previously identified to host a variety of faint tidal features. The sample consists of $\sim$6400 galaxies with imaging from either DECaLS or CFHTLS, as well as spectroscopy from SDSS. Visual inspection was used to identify around 800 systems with faint tidal features \citepalias{2024MNRAS.534.1459G,2013ApJ...765...28A}. These galaxies were further split into four non-exclusive categories depending on the morphology of their tidal features: \textit{arms}, \textit{streams}, \textit{shells}, and \textit{diffuse}. Galaxies without tidal features were split into elliptical, spiral, face-on disc, and miscellaneous disc categories based on the \citetalias{2022MNRAS.509.3966W} morphological classifications for the DECaLS sample or whether \citetalias{2013ApJ...765...28A} identified them as being in the red sequence or blue cloud. Cross-matching this sample to the MPA-JHU \citep{2003MNRAS.341...33K, 2004MNRAS.351.1151B}, we limited the sample to redshifts in the range $z\in\left[0.025, 0.15\right]$ and stellar masses between $\log_{10}\left(M_\star/M_\odot\right)\in\left[9.0, 11.6\right]$. We then selected control galaxies without tidal features by ensuring the mass and redshift distributions matched those of the galaxies with features.

Using the \citetalias{2007MNRAS.381..543W} PCA catalogue, we identified the evolutionary phase of each galaxy. The post-starburst (PSB) and star-forming (SF) phases were identified via selection criteria from \citetalias{2018MNRAS.477.1708P}, \citetalias{2022MNRAS.516.4354W}, and \citetalias{2022MNRAS.517L..92E}. The remainder of the phases -- starbursts (SB), intermediate (SB/SF), green valley (GV), and quiescent (Q) -- were based on those provided in \citetalias{2007MNRAS.381..543W}. Similarly, we used the BPT optical emission line diagnostic \citep{1981PASP...93....5B, 2006MNRAS.372..961K} with line flux estimates from the MPA-JHU catalogue to identify those galaxies with AGN activity.

We found a remarkable difference between the tidal and non-tidal galaxies identified in both the SB and PSB phases, with each category of tidal feature having a significantly larger fraction of galaxies in these phases.  Overall, we found that tidal galaxies were 6.6$\pm$0.9 times more likely to be in an SB phase, and 19.6$\pm$5.0 times more likely to have then rapidly quenched this intense formation, being in the PSB phase. We also found that the fractions varied depending on the type of tidal feature considered. For example, galaxies with \textit{shell} features were more likely to be quiescent than any of the other tidal galaxies and, similarly, less likely to be SB. On the other hand, \textit{arm} features were more likely to be in an SB phase and less likely to be PSB. We argued that these trends are qualitatively consistent with the likely merger processes involved. Indeed, they provide some of the first empirical evidence that the strength of merger-driven star formation and quenching processes in galaxies has a dependence on the detailed nature of the interaction, as traced by the tidal feature morphology.

For the AGN fraction, we found no significant difference between tidal and non-tidal galaxies, or between the subdivisions of tidal galaxies. This could be due to several factors, such as minor mergers being unable to produce sufficient accretion onto the central black hole, observing the AGN when it was not active, or simply that some galaxies have AGN, but they are obscured in the optical. Future works could attempt to determine if selecting AGN based on X-ray or IR data changes the result for these galaxies. An alternative reason that could explain the discrepancy between our work and other studies that do find an excess of AGN in post-merger galaxies could be the way in which we chose to normalise our AGN fraction, where we have excluded those galaxies that did not pass the SNR cut.  

\section*{Acknowledgements}
AJG is supported by a UK Science and Technology Facilities Council (UK STFC) studentship. AMNF is supported by UK Research and Innovation (UKRI) under the UK government's Horizon Europe funding guarantee [grant number EP/Z534353/1] and by the UK STFC [grant number ST/Y001281/1].

We thank Cassandra Barlow-Hall, James Aird, Sara Ellison, Jessica Howell and Thomas Stanton for their insightful and supportive discussions on the topic.

This project was made possible with the following \textsc{python} packages and software:
\textsc{astropy} \citep{2022ApJ...935..167A},
\textsc{astroquery} \citep{2019AJ....157...98G},
\textsc{github copilot} \citep{githubcopilot} for code auto-complete suggestions, 
\textsc{imageio} \citep{almar_klein_2024_14234213},
\textsc{matplotlib} \citep{2007CSE.....9...90H},
\textsc{numpy} \citep{harris2020array},
\textsc{pandas} \citep{mckinney-proc-scipy-2010},
\textsc{scipy} \citep{2020NatMe..17..261V},
\textsc{seaborn} \citep{Waskom2021},
and \textsc{skyproj} \citep{skyproj}.

This work used images from the Dark Energy Camera Legacy Survey. The Legacy Surveys consist of three individual and complementary projects: the Dark Energy Camera Legacy Survey (DECaLS; Proposal ID \#2014B-0404; PIs: David Schlegel and Arjun Dey), the Beijing-Arizona Sky Survey (BASS; NOAO Prop. ID \#2015A-0801; PIs: Zhou Xu and Xiaohui Fan), and the Mayall z-band Legacy Survey (MzLS; Prop. ID \#2016A-0453; PI: Arjun Dey). DECaLS, BASS and MzLS together include data obtained, respectively, at the Blanco telescope, Cerro Tololo Inter-American Observatory, NSF's NOIRLab; the Bok telescope, Steward Observatory, University of Arizona; and the Mayall telescope, Kitt Peak National Observatory, NOIRLab. Pipeline processing and analyses of the data were supported by NOIRLab and the Lawrence Berkeley National Laboratory (LBNL). The Legacy Surveys project is honored to be permitted to conduct astronomical research on Iolkam Du'ag (Kitt Peak), a mountain with particular significance to the Tohono O'odham Nation.

NOIRLab is operated by the Association of Universities for Research in Astronomy (AURA) under a cooperative agreement with the National Science Foundation. LBNL is managed by the Regents of the University of California under contract to the U.S. Department of Energy.

This project used data obtained with the Dark Energy Camera (DECam), which was constructed by the Dark Energy Survey (DES) collaboration. Funding for the DES Projects has been provided by the U.S. Department of Energy, the U.S. National Science Foundation, the Ministry of Science and Education of Spain, the Science and Technology Facilities Council of the United Kingdom, the Higher Education Funding Council for England, the National Center for Supercomputing Applications at the University of Illinois at Urbana-Champaign, the Kavli Institute of Cosmological Physics at the University of Chicago, Center for Cosmology and Astro-Particle Physics at the Ohio State University, the Mitchell Institute for Fundamental Physics and Astronomy at Texas A\&M University, Financiadora de Estudos e Projetos, Fundacao Carlos Chagas Filho de Amparo, Financiadora de Estudos e Projetos, Fundacao Carlos Chagas Filho de Amparo a Pesquisa do Estado do Rio de Janeiro, Conselho Nacional de Desenvolvimento Cientifico e Tecnologico and the Ministerio da Ciencia, Tecnologia e Inovacao, the Deutsche Forschungsgemeinschaft and the Collaborating Institutions in the Dark Energy Survey. The Collaborating Institutions are Argonne National Laboratory, the University of California at Santa Cruz, the University of Cambridge, Centro de Investigaciones Energeticas, Medioambientales y Tecnologicas-Madrid, the University of Chicago, University College London, the DES-Brazil Consortium, the University of Edinburgh, the Eidgenossische Technische Hochschule (ETH) Zurich, Fermi National Accelerator Laboratory, the University of Illinois at Urbana-Champaign, the Institut de Ciencies de l'Espai (IEEC/CSIC), the Institut de Fisica d'Altes Energies, Lawrence Berkeley National Laboratory, the Ludwig Maximilians Universitat Munchen and the associated Excellence Cluster Universe, the University of Michigan, NSF's NOIRLab, the University of Nottingham, the Ohio State University, the University of Pennsylvania, the University of Portsmouth, SLAC National Accelerator Laboratory, Stanford University, the University of Sussex, and Texas A\&M University.

BASS is a key project of the Telescope Access Program (TAP), which has been funded by the National Astronomical Observatories of China, the Chinese Academy of Sciences (the Strategic Priority Research Program "The Emergence of Cosmological Structures" Grant \# XDB09000000), and the Special Fund for Astronomy from the Ministry of Finance. The BASS is also supported by the External Cooperation Program of Chinese Academy of Sciences (Grant \# 114A11KYSB20160057), and Chinese National Natural Science Foundation (Grant \# 12120101003, \# 11433005).

The Legacy Survey team makes use of data products from the Near-Earth Object Wide-field Infrared Survey Explorer (NEOWISE), which is a project of the Jet Propulsion Laboratory/California Institute of Technology. NEOWISE is funded by the National Aeronautics and Space Administration.

The Legacy Surveys imaging of the DESI footprint is supported by the Director, Office of Science, Office of High Energy Physics of the U.S. Department of Energy under Contract No. DE-AC02-05CH1123, by the National Energy Research Scientific Computing Center, a DOE Office of Science User Facility under the same contract; and by the U.S. National Science Foundation, Division of Astronomical Sciences under Contract No. AST-0950945 to NOAO.

This work also makes use of data from the Sloan Digital Sky Survey III. Funding for SDSS-III has been provided by the Alfred P. Sloan Foundation, the Participating Institutions, the National Science Foundation, and the U.S. Department of Energy Office of Science. The SDSS-III web site is \url{http://www.sdss3.org/}.

SDSS-III is managed by the Astrophysical Research Consortium for the Participating Institutions of the SDSS-III Collaboration including the University of Arizona, the Brazilian Participation Group, Brookhaven National Laboratory, Carnegie Mellon University, University of Florida, the French Participation Group, the German Participation Group, Harvard University, the Instituto de Astrofisica de Canarias, the Michigan State/Notre Dame/JINA Participation Group, Johns Hopkins University, Lawrence Berkeley National Laboratory, Max Planck Institute for Astrophysics, Max Planck Institute for Extraterrestrial Physics, New Mexico State University, New York University, Ohio State University, Pennsylvania State University, University of Portsmouth, Princeton University, the Spanish Participation Group, University of Tokyo, University of Utah, Vanderbilt University, University of Virginia, University of Washington, and Yale University.
 
\section*{Data Availability}
This project used numerous publicly available datasets available online. The galaxies classified by \citet{2013ApJ...765...28A} are available at \url{https://content.cld.iop.org/journals/0004-637X/765/1/28/revision1/apj459673t4_mrt.txt}. Similarly, the \textit{GZD} galaxies identified by \citet{2022MNRAS.509.3966W} are available at \url{https://zenodo.org/records/4573248}. Line fluxes, stellar masses, and other galaxy parameters were taken from the \citet{2003MNRAS.341...33K} and \citet{2004MNRAS.351.1151B} MPA-JHU catalogue \url{https://wwwmpa.mpa-garching.mpg.de/SDSS/DR7/}. S{\'{e}}rsic indices for the galaxies were taken from the \citet{2011AJ....142...31B} NSA \url{https://www.sdss4.org/dr17/manga/manga-target-selection/nsa/}. Finally, the \citet{2007MNRAS.381..543W} PCA amplitudes were taken from \url{http://star-www.st-andrews.ac.uk/~web2vw8/downloads/DR7PCA.html}.

The data and code for this article will be shared upon a reasonable request to the corresponding author.



\bibliographystyle{mnras}
\bibliography{references} 

\appendix
\section{Alternate AGN fraction}\label{app:altagn}
In Section \ref{sec:bptresults} we presented the results of our analysis on the fraction of galaxies with AGN across the different tidal and non-tidal morphologies. In that section, we found no significant difference between the tidal and non-tidal categories, indicating that these merger events may not be significant drivers of AGN growth. However, we additionally claimed that when we included galaxies that failed the SNR cut in the computation of the fraction, we did indeed see an enhanced fraction of AGN in the tidal population by a factor of 2.1$\pm$0.3. For completeness and to aid in the comparison to other works, we present the results that led to that statement in this appendix. Again, we caution that we do not take this result as our main result as we cannot confidently say whether those that failed the cut were truly passive or active galaxies with poor spectra.

\begin{figure*}
    \includegraphics[width=\textwidth]{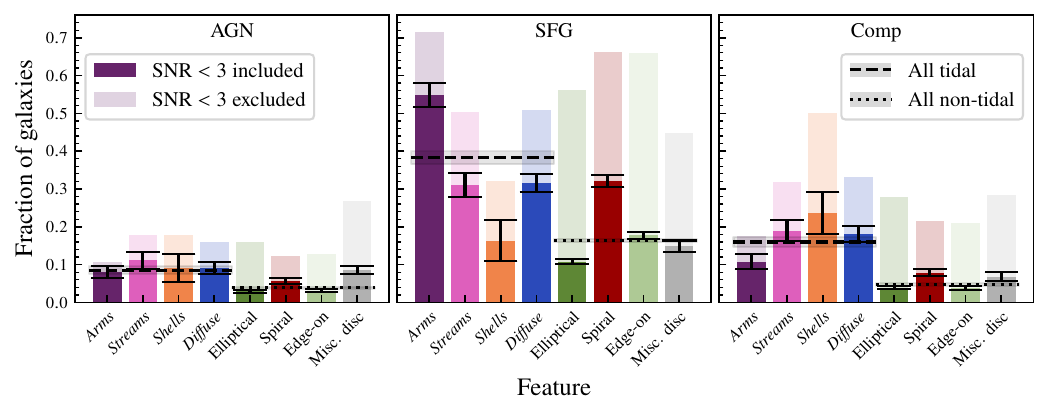}
    \caption{The fraction of each category of galaxy in each of the BPT classes but as a fraction of the total galaxy sample. Each bar shows the median fraction of that morphology that was identified to host AGN (left), be star-forming (centre), or a composite (Comp, right). The partially shaded bars indicate the same as a fraction of only those galaxies that passed the signal-to-noise cut, as they appear in Fig. \ref{fig:agn_fraction}. The dashed line indicates the all tidal fraction and, similarly, the dotted all non-tidal. The medians and errors were determined using the bootstrap method (see text).}
    \label{fig:altagn}
\end{figure*}

Figure \ref{fig:altagn} provides the fraction of galaxies in the sample identified as AGN, SFG, or Comp, normalised by the total galaxy sample (i.e. including those that failed the SNR cut). The partially shaded bars show the fractions as they were previously in Fig. \ref{fig:agn_fraction}. Indeed, when normalising in this way, there was an enhanced fraction of AGN in the tidal sample, with the tidal sample being 2.1$\pm$0.3 times more likely to host AGN.

This result is more similar to studies in the literature that do find an excess of AGN in merging and post-merger galaxies. In particular, it matches well to the excess of optically selected AGN in \citet{2023ApJ...944..168L}, who found that the excess of AGN in post-mergers was roughly a factor of $\sim$2. Additionally, \citet{2025A&A..L} found that 8 to 9 per cent of their merging galaxies from \textit{Euclid} hosted optical AGN, with an excess factor between 3 and 4, although they consider higher redshift galaxies ($z \in [0.5, 2.0]$). We recovered a similar result with 9.0$\pm$1.0 per cent of our tidal galaxies hosting AGN, although our excess was smaller.

In terms of the individual morphologies, there appeared to be a slight difference between some of the tidal classes, with \textit{streams} having a marginally higher frequency and \textit{arms} marginally lower. However, as before, we note that this was not a statistically significant result as the error bars accounted for the variation. Therefore, we reassert our claim that the AGN fraction does not appear to have any particular dependence on the morphology of the feature.

\bsp	
\label{lastpage}
\end{document}